\documentclass[letter]{IEEEtran}

\usepackage[utf8]{inputenc} 
\usepackage[T1]{fontenc}
\usepackage{url}              %
\usepackage{cite}             %

\usepackage[cmex10]{amsmath}  %
\usepackage{mleftright}       %
\mleftright                   %

\usepackage{graphicx}         %
\usepackage{booktabs}         %

\usepackage{algorithmicx}    %

\usepackage{amssymb,mathtools,bm,tikz,pgfplots,amsthm}
\pgfplotsset{compat=1.17}
\mathtoolsset{showonlyrefs}
\usepackage{enumerate}
\usepackage{scalerel}
\newcommand{\bplus}{\mskip2mu\scaleobj{0.85}{\boxplus}\mskip2mu}
\allowdisplaybreaks

\newtheorem{theorem}{Theorem}
\newtheorem{cor}[theorem]{Corollary}
\newtheorem{lemma}[theorem]{Lemma}

\theoremstyle{definition}
\newtheorem{definition}{Definition}
\newtheorem{remark}{Remark}
\newtheorem{assumption}{Condition}

\renewcommand{\mid}{\,|\,}

\newcommand{\bX}{\bm{X}}
\newcommand{\bY}{\bm{Y}}

\newcommand{\bc}{\bm{c}}

\newcommand{\be}{\bm{e}}

\newcommand{\bt}{\bm{t}}

\newcommand{\bx}{\bm{x}}

\newcommand{\q}{q}

\newcommand{\cA}{\mathcal{A}}
\newcommand{\cB}{\mathcal{B}}
\newcommand{\cC}{\mathcal{C}}

\newcommand{\cE}{\mathcal{E}}
\newcommand{\cF}{\mathcal{F}}
\newcommand{\cG}{\mathcal{G}}

\newcommand{\cM}{\mathcal{M}}

\newcommand{\cX}{\mathcal{X}}
\newcommand{\cY}{\mathcal{Y}}

\newcommand{\eps}{\epsilon}

\DeclareMathOperator{\diag}{diag}
\DeclareMathOperator{\gtr}{Tr}

\newcommand{\gx}{\bplus}
\newcommand{\one}{\boldsymbol{1}}

\newcommand{\dexit}{\delta}
\newcommand{\kexit}{Q}

\newcommand{\sym}[1]{\mathbb{S}_{#1}}
\newcommand{\simp}[1]{\Delta_{#1}}
\newcommand{\ser}{\mathrm{SER}}
\newcommand{\cch}{V}

\newcommand{\bEx}{\ensuremath{\mathbb{E}}}

\newcommand{\ex}[1]{\ensuremath{\mathbb{E}\left[ #1\right]}}
\newcommand{\pr}[1]{\ensuremath{\mathbb{P}\left[ #1\right]}}

\definecolor{henry}{rgb}{0,0.65,0}

\newif\ifarxiv

\arxivtrue

\begin{document}

\title{Achieving Capacity on Non-Binary Channels with \\ Generalized Reed--Muller Codes}

\IEEEoverridecommandlockouts

\author{%
   \IEEEauthorblockN{Galen Reeves\IEEEauthorrefmark{1}\IEEEauthorrefmark{2} and Henry D. Pfister\IEEEauthorrefmark{1}\IEEEauthorrefmark{3} \thanks{This research was supported in part by %
   NSF Grants 2106213, 2212437, and 1750362. Any opinions, findings, and conclusions or recommendations expressed in this material are those of the author and do not necessarily reflect the views of the National Science Foundation.}}
   
   \IEEEauthorblockA{Departments of Electrical and Computer Engineering\IEEEauthorrefmark{1},  Statistical Science\IEEEauthorrefmark{2}, and Mathematics\IEEEauthorrefmark{3} \\ Duke University}
}

\maketitle

\begin{abstract}
  Recently, the authors showed that Reed--Muller (RM) codes achieve capacity on binary memoryless symmetric (BMS) channels with respect to bit error rate.
  This paper extends that work by showing that RM codes defined on non-binary fields, known as generalized RM codes, achieve capacity on sufficiently symmetric non-binary channels with respect to symbol error rate.
  The new proof also simplifies the previous approach (for BMS channels) in a variety of ways that may be of independent interest.
\end{abstract}


\section{Introduction}

Generalized Reed--Muller (GRM) codes were introduced by Kasami, Lin, and Peterson in 1968~\cite{Kasami-it68} as the natural generalization of binary Reed--Muller (RM) codes~\cite{Muller-ire54,Reed-ire54} to non-binary alphabets.
GRM codes are closely related to other interesting code families including Reed--Solomon codes~\cite{Reed-jsim60}, multiplicity codes~\cite{Kopparty-stoc11}, and lifted codes~\cite{Guo-itcs13}.
These families remain interesting subjects of research due to their connections with topics such as local decodability and list decoding.

In 2016, it was established that sequences of RM codes can achieve capacity on the binary erasure channel (BEC)~\cite{Kudekar-stoc16,Kudekar-it17}.
This was followed by some extensions and related work~\cite{Sberlo-soda20,Abbe-it20,hkazla-stoc21}.
A nice tutorial overview of RM codes and results until 2020 is provided by~\cite{Abbe-it20a}.
Then, in 2021, the authors showed that RM codes achieve capacity on binary memoryless symmetric (BMS) channels with respect to bit error rate~\cite{Reeves-arxiv21}.

The main result of this paper is the following theorem.
We note that all terminology will be defined in later sections. 

\begin{theorem} \label{thm:main}
Consider a memoryless channel $W$ with capacity $C$ whose input alphabet is $\cX=\mathbb{F}_q$ and let $G$ be the symmetry group of the channel. Suppose one of the following holds: 
\begin{enumerate}[(i)]
\item $G$ contains the affine group over $\mathbb{F}_q$;
\item  $q$ is prime, $G$ contains the additive group of $\mathbb{F}_q$, and the smallest principal inertia component of $W$ (for the uniform input distribution) is strictly positive. 
\end{enumerate}
Then, for every sequence of GRM codes over $\mathbb{F}_q$ with strictly increasing blocklength and rate converging to $R \in [0,C)$, the symbol-error rate (SER) under symbol-MAP decoding converges to zero.
\end{theorem}

\begin{cor} \label{cor:prime}
If $q$ is prime and the channel symmetry group is transitive (i.e., the channel is symmetric) but it does not contain the additive group of $\mathbb{F}_q$, then the input alphabet can be relabeled so that the conclusion of Theorem 1 still holds.
\end{cor}

\begin{cor} \label{cor:main}
Consider a memoryless channel $W$ with input alphabet $\cX=\mathbb{F}_q$ and let $I_s$ be the mutual information between its input and output with a uniform input distribution.
Consider block-coded transmission using group symmetrization over the affine group of $\mathbb{F}_{\q}$ (i.e., each channel use is modulated by a random affine map known at the receiver).
Then, for every sequence of GRM codes over $\mathbb{F}_q$ with strictly increasing blocklength and rate converging to $R \in [0,I_s)$, the SER under symbol-MAP decoding converges to zero.
\end{cor}

For the purposes of our analysis, there are two finite-input channels of interest. The first is the memoryless channel $W$ (with capacity $C$) over which the codeword $\bX$ is transmitted and the output $\bY$ is received. Our goal is to show that code rates strictly less than $C$ can be achieved with vanishing symbol error rate.  %
This involves analyzing a second channel from $X_0$ to $Y_{\sim 0}$, which we call  the \emph{coset} channel due to the group structure in the code. 

The high-level idea of our proof is to first show that, as the blocklength increases,  the sequence of coset channels converges to a \emph{deterministic} channel (i.e., a channel for which a minimal sufficient statistic is a non-random function of its input).
For binary inputs, the only deterministic channels are the perfect channel and uninformative channel but, for non-binary inputs, there are other possibilities.
In the remainder of the proof, we use channel symmetry and the area theorem to argue that this limiting channel must be the perfect channel whenever the rate of the code is strictly less than capacity. 

The two conditions appearing in Theorem~\ref{thm:main} have different implications for the symmetry group of the implied coset channel.
Case (i) implies doubly transitive symmetry of the coset channel, which simplifies much of the analysis. 
Case (ii) implies transitive symmetry of the coset channel, and in this case, we need the  assumption that $q$ is prime and some additional arguments to rule out the possibility that the coset channel converges to a deterministic limit that is neither perfect nor uninformative.

\subsubsection*{Comparison with prior work} 
The proof for binary RM codes on BMS channels~\cite{Reeves-arxiv21} is based on the convergence of the power series expansion of the binary entropy function around the uninformative point.
This fails for the non-binary case because the analogous power series converges only on a small subset of the domain.
Instead, an approach inspired by strong data-processing inequalities  (Lemma~\ref{lem:strong_concavity}) is used here to bound mean-squared error in terms of mutual information.

 In \cite{Reeves-arxiv21}, the ``influences'' of two subsets (of channel outputs) on the conditional mean estimator are bounded separately using two different arguments.  In this paper, all influences are bounded using a single simpler argument (Lemma~\ref{lem:rate_cover}) that uses extrinsic information transfer (EXIT) functions~\cite{Ashikhmin-it04} rather than generalized EXIT functions~\cite{Measson-it09}.

Our analysis of channels builds on the framework developed by Blackwell~\cite{Blackwell-ams53} and Le Cam~\cite{lecam:1986}.
Our introduction and analysis of the overlap matrix is related to principal inertial components (PICs)~\cite{witsenhausen:1975,calmon:2017} and strong data-processing inequalities~\cite{Makur-it18,Raginsky-isit11,Raginsky-isit13}.
A key innovation in this work is Lemma~\ref{lem:rate_cover}, which combines these ideas with a differential analysis of channels enabled by the EXIT area theorem.

\ifarxiv
After the conference version of this paper~\cite{Reeves-isit23} was accepted, a proof that RM codes have vanishing block-error rate on BMS channels, at any rate less than capacity, was posted to arXiv~\cite{Abbe-arxiv23}. 
It would be interesting to see if the methods in~\cite{Abbe-arxiv23} can be combined with this work to establish vanishing block error rate in the non-binary setting.
\fi

\subsubsection*{Notation} 
The real numbers and are denoted by $\mathbb{R}$, the natural numbers are denoted by $\mathbb{N} \coloneqq \{1,2,\ldots\}$, and $\mathbb{N}_0 \coloneqq  \mathbb{N} \cup \{0\}$.
For $N \in \mathbb{N}_0$, a range of natural numbers is denoted by $[N] \coloneqq \{0,1,\ldots,N-1\}$. 
Let $\simp{q}$ denote the probability simplex on $q$ elements and $\be_{i} \in \mathbb{R}_q$ be the $i$-th standard basis vector.
We use $\mathbb{F}_q$ to denote the Galois field with $q$ elements. %
For a set $\cX$, the $N$-element vector $\bm{x} \in \mathcal{X}^N$ is denoted by boldface and is indexed from 0 so that $\bm{x}=(x_0,,\ldots,x_{N-1})$.
For an $M$-element index set $A=\{a_0,a_1,\ldots,a_{M-1}\}\subseteq [N]$ with $a_0<a_1<\cdots<a_{M-1}$, we define the subvector $x_A = (x_{a_0},x_{a_1},\ldots,x_{a_{M-1}}) \in \cX^M$ without using boldface. A single random variable is denoted by a capital letter (e.g., $X,Y,Z$). Vectors of random variables are denoted by boldface capital letters (e.g., $\bm{X},\bm{Y},\bm{Z}$).
All unspecified logarithms (i.e., $\log$'s) are taken base-$\q$ and thus expressions involving entropy and mutual information are reported in \emph{qits}.

\ifarxiv
This is an extended version of the paper~\cite{Reeves-isit23}.
\fi

\section{Channels} 
\label{sec:sym_channels}

We assume throughout that  $q \in \mathbb{N}$ with $q \ge 2$ and $\cX =  \{0,1,\dots, q-1\}$. The symmetric group  $\sym{q}$ is the set of bijective functions (i.e., permutations) mapping $\cX$ to  $\cX$ with the group operation given by composition. A permutation group $G \subseteq \sym{q}$ is \emph{transitive} if, for each $x,x' \in \cX$, there exists a permutation in $G$ the maps $x$ to $x'$.
Likewise, it is \emph{doubly transitive} if, for any $x_1,x_2, x_1',x_2' \in \cX$ with $x_1 \ne x_2$ and $x_1' \ne x_2'$, there exists a permutation in $G$ that maps $x_k$ to $x_k'$ for $k =1,2$. %

We define the action of the symmetric group $\sym{q}$ on $\mathbb{R}^q$ (and the probability simplex $\Delta_q \subseteq \mathbb{R}^q$) according to \vspace{-0.75mm}
\begin{align}
     \sigma (v_0, \dots, v_{q-1}) =   (v_{\sigma (0)}, \dots, v_{\sigma(q-1)}) \vspace*{-1.25mm}
\end{align}
for every $ \sigma \in \sym{q}$ and $ (v_0, \dots, v_{q-1}) \in \mathbb{R}^q$. This operation is extended to a probability measure $P$ on $\mathbb{R}^q$ via the pushforward measure $\sigma P$ defined by \vspace{-0.75mm}
\begin{align}
    (\sigma P)(B) = P( \sigma^{-1} B) , \quad \forall B \in \cB \vspace*{-1.25mm}
\end{align}
where $\sigma^{-1} B = \{ \sigma^{-1} p \mid p \in B \} $.

\subsection{Finite-Input Channels} \label{sec:channels}

A $\q$-ary input channel $W$ is a conditional distribution mapping from an input $x \in \cX$ to a probability measure $W( \cdot \mid x)$ on a measurable space $(\cY, \cA)$. When convenient, we use the compact notation $W_x(\cdot) = W(\cdot \mid x)$. %

Following the approach of Blackwell~\cite{Blackwell-ams53}, we introduce a standard version of the channel whose output alphabet is the probability simplex.  For a channel $W$,  the  \emph{canonical map} $\phi \colon \cY \to \simp{\q}$ is defined by
\begin{align}
    \phi (y) &\coloneqq \big( \phi_0(y), \phi_1 (y), \ldots, \phi_{\q-1} (y) \big) ,
\end{align}
where  $\phi_x (y) \coloneqq (d W_x / d \bar{W})(y)$ is 
the Radon-Nikodym derivative of $W_x$ with respect to the reference measure 
\begin{align} \label{eq:wbar}
    \bar{W}(A) &\coloneqq \sum_{x \in \cX} W( A\mid x), \quad A \in \cA. 
\end{align}
The canonical map can also be viewed as  the posterior pmf of with respect to a uniform prior distribution, i.e., $\phi_x(y)$ is the probability that the input is $x$ given the output is $y$.

Composing the channel $W$ with its canonical map produces a new channel, $W^\mathrm{s}$,  on the output space $(\simp{q}, \cB)$ satisfying
\begin{align}
    W^{\mathrm{s}}(B\mid x) = W(\phi^{-1} B \mid x) , \quad \forall x \in \cX, \forall B \in \cB. 
\end{align}
Because the canonical map is a sufficient statistic for the channel input, the mapping from $W$ to $W^\mathrm{s}$ preserves the relevant properties of the channel,  such as its capacity and minimum error probability.

Following Blackwell's definition of a standard experiment~\cite{Blackwell-ams53}, we call a channel \emph{standard} if its canonical map is the identity map, and we refer to $W^\mathrm{s}$ as the standard channel associated with $W$. 
Furthermore, we will call two channels \emph{Blackwell equivalent} if they have the same standard channel.

\subsection{Channel Symmetry}

In communication theory, the term symmetric channel is used to refer to a variety of related (but distinct) symmetry conditions~\cite{Polyanskiy-it13,Slepian-bell56}.
This paper uses the following definition due to its compatibility with Blackwell equivalence.

\begin{definition}[Channel Symmetry Group]
The symmetry group $G$ of a $\q$-ary channel $W$ is the permutation group
\begin{align*}
G &=
\{ \sigma \in \sym{q} \mid \forall x \in  \cX, \forall B\in \cB,     W^{\mathrm{s}} (\sigma B \mid \sigma x)  = W^{\mathrm{s}} (B \mid x)\},
\end{align*}
where $W^{\mathrm{s}}$ is the standard channel associated with $W$. In other words, $G$ is the group of all permutations $\sigma$ such that the distribution of the canonical map $\phi(Y)$ under input $\sigma x$ is equal to the distribution of $\sigma \psi(Y)$ under input $x$.
\ifarxiv
This can also be written compactly as $G=\{ \sigma \in \sym{q} \mid \forall x \in  \cX,  \;   \sigma W_{x}^{\mathrm{s}}=  W^{\mathrm{s}}_{\sigma x}\}$.
\fi
\end{definition}

A channel is called \emph{symmetric} if its symmetry group  is transitive. Its well known that this condition is sufficient to ensure that the capacity of the channel is achieved by the uniform input distribution~\cite{Polyanskiy-it13}. More generally, for any decision-theoretic problem whose loss function has the same symmetries as the channel,  the uniform input distribution maximizes the expected loss~\cite[Chapter~6]{eaton:1989}.

A slightly stronger notion of symmetry occurs when the symmetry group of a channel is associated with a group structure on the input alphabet. %
Let  $(\cX, \gx)$ be a group with binary operation denoted by $\gx$, and assume without loss of generality that $0$ is the identity element. Each $x \in \cX$ defines a permutation $\sigma_x \in \sym{q}$ according to $\sigma_x x' = x \gx x' $ for every $x' \in \cX$.   By Cayley's theorem, the group $(\cX, \gx)$ is isomorphic to the permutation group $H \coloneqq \{\sigma_x  \mid x \in \cX\}$, which we will refer to as the permutation representation of $(\cX, \gx)$. 
The channel $W$ is called \emph{group symmetric} if its symmetry group contains $H$ as a subgroup.

While there exist channels that are symmetric but not group symmetric, this can occur only if $q$ is not a prime power.

\begin{lemma} \label{lem:sylow_group}
If a channel $W$ is symmetric with $q$ equal to a prime power, then it is group symmetric.  
\end{lemma}

\subsection{Group Symmetrization}
\label{sec:group_sym}

There is a simple (and commonly used) process that can equip any channel with any desired symmetry.
Moreover, if the channel capacity is achieved by a uniform input distribution, then this process does not change the capacity.
Let $W$ be a channel with input alphabet $\cX$ and let $H$ be any subgroup of $\sym{\q}$.
Now, define a new channel $W'$ that, for the input $x\in \cX$, chooses a uniform random element of $\sigma \in H$ and transmits $\sigma x$ through $W$.
Then, the output of $W'$ is defined to be $(Y,\sigma)$ where $Y\in \cY$ is the output of $W$.
We call this operation group symmetrization and the symmetry group of the resulting channel must contain $H$ as a subgroup.
The standard channel $W'\,\!^{\mathrm{s}}$ satisfies
\[
W'\,\!^{\mathrm{s}} ( B \mid x) = \frac{1}{|H|} \sum_{\sigma \in H} W^{\mathrm{s}} (\sigma^{-1} B \mid x). 
\]

If $H$ is transitive, then the group symmetrization operation implies that the effective input distribution seen by the original channel $W$ is uniform.
So, if one is content to include group symmetrization in the system, then any desired channel symmetry can be engineered while still achieving the rate $I_s$ equal to the mutual information between the channel input and output under a uniform input distribution.

\subsection{Overlap Matrix}
\label{sec:overlap}

We define \emph{overlap matrix} $Q \in \mathbb{R}^{q\times q}$ associated with a $q$-ary channel $W$ according to 
\begin{align}
\kexit_{x,x'} \coloneqq   \int \phi_{x} (y) \, \phi_{x'}(y) \bar{W}(dy). 
\end{align}
 This matrix is symmetric, positive semidefinite, and doubly stochastic. Thus, its eigenvalues are real positive numbers that satisfy $1  = \lambda_0 \ge \lambda_1 \ge \dots \ge \lambda_{\q-1} \ge 0$. 
It can be verified that the eigenvalues of index $x\ge 1$ correspond to the principal inertia components (PICs) of the channel with respect to the uniform input distribution \cite{witsenhausen:1975,calmon:2017}. 
The smallest PIC $\lambda_{q-1}$, which plays a prominent role in our analysis, has been considered previously in the context of perfect privacy~\cite{calmon:2017}.

For random variables $(X,Y)$,  the  \emph{symbol error rate} (SER) is defined by
\begin{align}
\ser(X\mid Y) \coloneqq \min_{X - Y - \hat{X}} \pr{ X \ne \hat{X}},
\end{align}
where the minimum is over all Markov chains $X - Y - \hat{X}$.

\begin{lemma}\label{lem:SERtoOverlap}
For any input-output pair $(X,Y)$ through a $q$-ary channel with overlap matrix $Q$, we have
\begin{align}
\ser(X\mid Y) & \le 1- \gtr( \diag(p) Q ), 
\end{align}
where $p \in \Delta_q$ is the prior pmf of $X$.
\end{lemma}

The following lemma will also us to connect MMSE estimation error with conditional mutual information.

\begin{lemma}%
\label{lem:strong_concavity}%
For any  Markov chain $S- T  - X - Y$ where  $Y$ is an observation of $X$ through $q$-ary channel, we have,%
\begin{align*}
\MoveEqLeft I(X ; Y \mid S) - I(X ; Y \mid T) \\
&
\ge \frac{\lambda^2_{q-1}}{2 \ln q} \ex{ \| \ex{ \be_{X} \mid T} - \ex{ \be_{X} \mid S}  \|^2}, 
\end{align*}
where $\lambda_{q-1}$ is the minimal eigenvalue of the overlap matrix
\end{lemma}

For a $q$-ary channel $W$ with canonical map $\phi$, we define the squared-error discrepancy 
\begin{align}
\dexit  \coloneqq \ex{ \| \phi(Y ) - \ex{ \phi(Y ) \mid X}\|^2}  \label{eq:dexit}
\end{align}
where $Y$ is the output for a uniformly distributed input $X$. Note that $\delta$ is zero if and only if the output of the standard channel is determined uniquely by the input (i.e., the channel is deterministic). 
This discrepancy can also be expressed in terms of the overlap matrix $Q$ or the PICs:
\begin{align}
\dexit &=  \frac{1}{q} \gtr(\kexit)- \frac{1}{q} \gtr( \kexit^2)  = \frac{1}{\q} \sum_{x=1}^{\q-1} \lambda_x (1- \lambda_x). \label{eq:dexit_alt}
\end{align}
If $\dexit$ is close to zero, then the PICs are clustered near the boundaries $0$ and $1$ of the unit interval. The next result gives sufficient conditions under which the  PICs are all close to the same boundary point.%

\begin{lemma} \label{lem:Bconstraints} 
Consider a $q$-ary channel with symmetry group $G$ and squared-error discrepancy $\dexit$. Suppose one of the following  holds:
\begin{enumerate}[(i)]
    \item $G$ is doubly transitive; 
    \item $G$ is transitive, $q$ is prime, and $4 q^2 \dexit < 1$. 
\end{enumerate}
Then, the overlap matrix $\kexit$ satisfies
\begin{align}
\min_{b \in \{1,q\} } | \gtr(\kexit) - b| \le  2 \q \dexit. \label{eq:trQtob}
\end{align}
\end{lemma}

\ifarxiv
\begin{remark}
To see that the conditions in Lemma~\ref{lem:Bconstraints} are not superfluous, consider the channel with input $\cX = \{0,1,2,3\}$ and output $Y = X+N \pmod{4}$ where $N$ is uniform on $\{0,2\}$. This channel is deterministic in the sense that its canonical map is determined uniquely by the input:%
\begin{align}
\phi(Y) = \frac{1}{2} \be_{X} + \frac{1}{2} \be_{X+2}
\end{align}
Hence, the squared-error discrepancy is $\delta = 0$. 
The channel is also symmetric because its symmetry group $G$ contains all cyclic shifts. For example, adding an element $x' \in \cX$ to the input before transmitting is equivalent to adding the same element to the output $Y$.
However, the channel does not satisfy the conditions of Lemma~\ref{lem:Bconstraints} because $G$ is not doubly transitive and $q$ is not prime. The  overlap matrix satisfies $\gtr(Q) = 2$, which violates~\eqref{eq:trQtob}.
\end{remark}
\else
\fi

\section{Codes}
\label{sec:codes}

\subsection{Group Codes}

In coding theory, the term group code is used to refer to a few related (but distinct) mathematical objects.
These include binary codes closed under modulo-2 addition~\cite{Slepian-bell56}, sets of points real space generated by a group of orthogonal transformations applied to a single point~\cite{Slepian-bell68}, and codes whose codewords are elements of a group ring~\cite{Berman-cyber67}.
The group structure of binary codes was recognized early and exploited in~\cite{Slepian-bell56,Fontaine-ire59}.
Later, similar ideas were developed for non-binary codes and channels~\cite{Slepian-bell68,Forney-it91}.
In this paper, we use the following definition.

\begin{definition}[Group code]
Let $(\cX,\gx)$ be a group where $0\in \cX$ is the identity.  A set $\cC \subseteq \cX^N$ is called a \emph{group code over $(\cX, \gx)$} if %
the set $\cC$ forms a group with respect to the binary operation $
\bx \gx \bx' = (x_0 \gx x_0',x_1 \gx x'_1, \dots, x_{N-1} \gx x'_{N-1})$.
\end{definition}

\begin{definition}[Matched to  channel]
A group code over $(\cX,\gx)$ is matched to channel $W$ if the symmetry group of $W$ contains the permutation representation of $(\cX, \gx)$ as a subgroup.  
\end{definition}

 For a group code matched to a channel, the code and channel together have a property which is akin to the geometric uniformity defined by Forney~\cite{Forney-it91}.
Similar ideas were explored more recently for group codes over integer rings~\cite{Como-it09}.
In particular, one gets the uniform error property where the error rate of the optimal decoder is independent of the transmitted codeword.

\subsection{The Coset Channel} 
\label{sec:coset}

For this section, assume that $\cC \subseteq \cX^N$ is a group code over $(\cX, \gx)$ and that
for all $x \in \cX$, there exists $\bc \in \cC$ such that $c_0 =x$.  Under these assumptions, the cosets of the  subgroup $\{ \bc \in \cC \mid c_0 = 0\} $ partition the code $\cC$ into $q$ sets of equal size. 
We define the \emph{coset channel} to be the channel from $X_0$ to $Y_{\sim 0}$ defined as follows:
\begin{enumerate}
    \item Given the input $x_0 \in \cX$, a codeword $\bX$ is drawn uniformly from the coset $\{ \bc \in \cC \mid c_0 = x_0\}$. 
    \item The output $Y_{\sim 0}$ is a memoryless observation of $X_{\sim 0}$ through the channel $W$. 
\end{enumerate}
From now on, we will use $V$ to denote the coset channel and $\psi$ to denote its canonical map. Composing $V$ with $\psi$ gives the standard coset channel $V^\mathrm{s}$.

\begin{lemma} \label{lem:coset_chan_sym}
If the code is matched to the channel $W$
and, for all $u \in \cX$, there exists $\bc \in \cC$ with $c_0 = u$, then 
the coset channel $V$ is group symmetric. Also, the output of the standard coset channel does not depend on which coset element is chosen, i.e., for an arbitrarily distributed input $X_0$, the output $\psi(Y_{\sim 0})$ is conditionally independent of $X_{\sim 0}$ given $X_0$. 
\end{lemma}

If the code has additional symmetries that are matched to the channel, then the symmetry group of the coset channel may be larger. For the following result, let $G$ be the permutation group of the channel $W$, let $H$ be the permutation representation of the group $(\cX, \gx)$, and let the group of homogeneous alphabet relabelings that preserve the code be given by
\[
F \coloneqq \left\{\sigma \in \sym{\q} \mid \forall \bm{c} \in \cC, \big (\sigma (c_0),\sigma (c_1),\ldots,\sigma (c_{N-1}) \big) \in \cC \right\}. 
\]
Since both $F$ and $G$ are subgroups of $\sym{q}$, their intersection $F' \coloneqq F \cap G$ is also a subgroup of $\sym{q}$. 

\begin{lemma} \label{lem:HF_matched}
If the code is matched to the channel $W$, then the symmetry group of the coset channel $V$ contains the group $\langle F', H  \rangle$ generated by $F'$ and $H$.  
\end{lemma}

For the case where $q$ is not prime, our proof technique requires that the coset channel has doubly transitive symmetry. In view of Lemma~\ref{lem:HF_matched}, a sufficient condition for this can be readily verified when $q$ is a prime power and $\cC$ is a linear code over $\cX=\mathbb{F}_q$ (and hence a group code with respect to the additive group of $\mathbb{F}_q$). Furthermore, if the code contains the all ones codeword (i.e., $(1,\dots, 1) \in \cC)$ then its group  of homogeneous alphabet relabelings contains the affine group  over $\mathbb{F}_q$, which is defined by $A_{\q} \coloneqq \{ \sigma_{a,b} \in \sym{q} \mid a \in \mathbb{F}_q \setminus \{0\}, b \in \mathbb{F}_q \}$ where $\sigma_{a,b} (x) = (a \cdot x) + b$ uses $\mathbb{F}_{q}$ addition and multiplication.

\begin{lemma} \label{lem:linear_code_affine}
If $\cC \subseteq \mathbb{F}^N_q$ is a linear code that contains the all ones codeword and the symmetry group of the channel $W$ contains the affine group $A_{q}$, then  the code is matched to the channel $W$ and the symmetry group of the coset channel $V$ is  doubly transitive. 
\end{lemma}

Lastly, we recall that the permutation automorphism group of the code of a $\cC \subseteq \cX^N$ is the group of permutations $\pi \in \sym{N}$ such that $(c_{\pi(0)}, c_{\pi(1)}, \dots, c_{\pi(N-1)} ) \in \cC$ for all $c \in \cC$. If this group is transitive then the coset channel defined for symbol position $0$ is Blackwell equivalent to the coset channel for any other position $i \in [N]$.

\subsection{Generalized Reed--Muller Codes and Puncturing}

The natural generalization of binary RM codes to non-binary alphabets was introduced by Kasami et al.\ in 1968 and dubbed Generalized Reed--Muller (GRM) codes~\cite{Kasami-it68}.
The GRM code RM$_q (r,m) \subseteq \mathbb{F}_q^N$ is a length $N=q^m$ linear code over $\mathbb{F}_q$.
Like binary RM codes, GRM codes can be defined as polynomial evaluation codes.
\ifarxiv
A more detailed description of GRM codes can be found in Appendix~\ref{sec:grm_codes}.
\fi

The rate of RM$_q (r,m)$,  denoted by $R_q (r,m)$, is  computed using the base-$q$ logarithm.
Thus, it equals the number of $\mathbb{F}_q$ information symbols (i.e., the dimension of the code) divided by the number of $\mathbb{F}_q$ codeword symbols. 

\ifarxiv
The following lemma bounds the change in rate caused by reducing $m$. The proof is in Appendix~\ref{sec:grm_codes}. 
\fi

\begin{lemma} \label{lem:grm_rate_diff}
For integers $q\geq 2$, $0 \leq r \leq m(q-1)$, and $0 \leq k < m-r$, the rates of RM$_q(r,m-k)$ and RM$_q(r,m)$ satisfy
\[ R_q (r,m-k) - R_q(r,m) \leq \frac{4k}{\sqrt{m-k}}. \]
\end{lemma}

\begin{definition}[Punctured Code] \label{def:punct_code}
For a code $\mathcal{C}\subseteq \mathbb{F}_q^N$ and index set $I\subseteq[N]$,  the punctured code formed by the symbol positions indexed by $I$ is given by $\cC_I \coloneqq \{\bm{c}_{I} \in \mathbb{F}_q^{|I|} \,\mid \,  \bc \in \cC\}$. 
\end{definition}

\begin{remark} 
Although  one may also consider a puncturing operation that includes reordering of code symbols, this is not needed for our results.
For our definition, the symbols are kept in the same order but their indices are renumbered. 
\end{remark}

\ifarxiv
The following lemma is proved in Appendix~\ref{sec:grm_codes}.
\fi

\begin{lemma}[GRM Puncturing] \label{lem:grm_punct}
If one punctures the code $\cC = \mathrm{RM}_q (r,m)$ by keeping only the first $q^{m-k}$ symbol positions (i.e., giving $\cC_I$ with $I=[q^{m-k}]$),
then $\cC_I = \mathrm{RM}_q (r,m-k)$.
Moreover, puncturing a uniform random codeword from $\cC$ gives a uniform random codeword from $\cC_I$.
\end{lemma}

\section{Main Results}
\label{sec:main}

\subsection{SER of the Coset Channel}
\label{sec:ser_coset}

This section gives bounds on the SER of the  coset channel (defined in Section~\ref{sec:coset}) under the following conditions: %

\begin{assumption}[Code]\label{assumption:code}
The input $\bX$ is distributed uniformly of the codewords of a $q$-ary group code $\cC$ that has code rate $R$. 
The code has a transitive permutation automorphism group and, for each $x\in \cX$, there exists $\bc \in \cC$ with $c_0 = x$.
\end{assumption}

\begin{assumption}[Channel]\label{assumption:channel}
The  output $\bY$ is an observation of the input through a symmetric memoryless channel $W$ that is matched to the code and has capacity $C$. 
\end{assumption}

Under these conditions, $X_0$ is uniformly distributed, the coset channel $V$ is group symmetric, %
and the SER of the coset channel is an upper bound on the maximal SER of the code:
\begin{align}
 \max_{i \in[N]}  \ser(X_i \mid \bY)  &= \ser(X_0 \mid \bY)  
  \le \ser(X_0 \mid Y_{\sim 0 })
\end{align}

For the purposes of analysis, we introduce a degraded family of channels that interpolates between $W$  and an uninformative channel. Specifically,  we define $W_t$ to be the composition of $W$ and an erasure channel with erasure probability $t \in [0,1]$. The implied coset channel is denoted by $V_t$ and its squared error discrepancy is given by %
\begin{align}
\delta(t)  & \coloneqq 
 \ex{ \| \Psi(t) - \ex{ \Psi(t) \mid X_0}\|^2 }, \quad 0 \le t \le 1
\end{align}
where $\Psi(t) \coloneqq \psi_t(Y_{\sim 0}(t)) = \ex{ \be_{X_0} \mid Y_{\sim 0}(t)}$. %
Here, the second expression for $\Psi(t)$  holds because $X_0$ is uniformly distributed, and thus the canonical map is equal to the posterior pmf. 
Finally, the \emph{average discrepancy} is defined to be
$$\dexit_\mathrm{avg} \coloneqq \int_0^1 \dexit(t) \, dt.
$$

We begin with a lower bound on the overlap matrix of the coset channel.  In combination with Lemma~\ref{lem:SERtoOverlap}, %
this bound shows that the SER of the coset channel is strictly less than the trivial upper bound $1 -1/q$ whenever the code rate $R$ is strictly less than the capacity of the channel $W_t$.%

\begin{lemma}%
\label{lem:weak_bound} 
Assume that Conditions~\ref{assumption:code} and \ref{assumption:channel} hold with $R < C$. For all 
$0\le t< 1-  R/C$, 
the overlap matrix $\kexit(t)$ of the coset channel $\cch_t$ satisfies 
\begin{align}
\gtr(\kexit(t))\ge q^{ C  -  R/(1-t) }. %
\end{align}
\end{lemma}

Next, we combine  Lemma~\ref{lem:weak_bound} with the constraints on the overlap matrix in Lemma~\ref{lem:Bconstraints} to provide a stronger bound on the SER in terms of the average discrepancy.

\begin{lemma}\label{lem:coset_ser}
Assume that Conditions~\ref{assumption:code} and \ref{assumption:channel} hold with $R< C$ and the average discrepancy satisfies 
\begin{align} \label{eq:delta_avg_assumption_alt}
\dexit_\mathrm{avg} \le  \frac{(1 - R/C) ( q^{\frac{1}{2} (C  -R) }-1) }{q} .
\end{align}
Further, suppose that one of the following conditions holds: 
\begin{enumerate}[(i)]
\item the symmetry group of $\cch$ is doubly transitive, or 
\item the symmetry group of $\cch$ is  transitive,  $q$ is prime, and \[
\dexit_\mathrm{avg} < \frac{1 - R/C}{ 8 q^2}.
\]
\end{enumerate}
Then, the SER satisfies
  \begin{align*}
\ser (X_0 \mid  Y_{\sim 0}) \le  \frac{4 \dexit_\mathrm{avg}}{1- R/C}. 
\end{align*}
\end{lemma}

Finally, we provide a link between the average discrepancy and the entropy rates of subsets of the code. The following result is obtained by combining a decomposition of the squared error discrepancy, via the Efron-Stein-Steele inequality,  with Lemma~\ref{lem:strong_concavity} and the EXIT area theorem. 

\begin{lemma} \label{lem:rate_cover}
Assume Conditions~\ref{assumption:code} and \ref{assumption:channel} hold. Furthermore, assume that overlap matrix of $W$ 
has minimal eigenvalue $\lambda_\mathrm{min} > 0 $.  
Let $\cB$ be a collection of subsets of $[N]$ such that
 \begin{enumerate}[(i)]
     \item $\bigcap_{B \in \cB} B = \{0\}$
     \item For each $B \in \cB$, the punctured code $\cC_B$ has a transitive permutation automorphism group.%
 \end{enumerate}
Then, we have
\begin{align}
\dexit_\mathrm{avg}&\le \frac{2 \ln q}{\lambda^2_\mathrm{min} } \,  \sum_{B \in \cB}\left( \frac{ H(X_B)}{ |B|} - R  \right).
\end{align}
\end{lemma}

\subsection{Proof of Theorem~\ref{thm:main}}%

We begin by verifying the conditions used in Section~\ref{sec:ser_coset}. Since the GRM code is a linear code over $\mathbb{F}_{\q}$, it is automatically a group code under the additive group of $\mathbb{F}_{\q}$. Also, is is well known that GRM codes have doubly transitive permutation automorphism groups~\cite{Kasami-it68}.
For a linear code, each code position either takes all possible values or is always 0.
Thus, for all $x \in \cX$, there is a $\bc \in \cC$ such that $c_0 = x$ because otherwise $\cC$ would only contain the all zero codeword due to transitive symmetry and have rate zero. Together, these results imply Condition~\ref{assumption:code} is satisfied.

Next, we note that cases \emph{(i)} and \emph{(ii)} of Theorem~\ref{thm:main} both require that $G$ contains the additive group of $\mathbb{F}_q$. This implies that the channel $W$ is group symmetric and the code is matched to the channel.
Thus, Condition~\ref{assumption:channel} is also satisfied.

Under these conditions, we can apply Lemma~\ref{lem:coset_chan_sym} to see that the coset channel is is group symmetric and, for case \emph{(ii)}, we can apply Lemma~\ref{lem:linear_code_affine} to see that it is doubly transitive. Having verified the assumptions of Lemma~\ref{lem:coset_ser}, we conclude that for any $\eps \in (0,1]$ there exists $\delta^* > 0$, such that if $ R_q(r,m) \le (1- \eps) C$ and $\delta_\mathrm{avg} \le \delta^*$, then %
\begin{align}
\ser (X_0 \mid Y_{\sim 0}) \le  \frac{4 \delta_\mathrm{avg}}{1- R/C}.
\end{align}

The final step of the proof  is to show that $\delta_{\mathrm{avg}}$ converges to zero as $m \to \infty$.  If $C=0$, the theorem is vacuous, so we assume $C>0$. If the channel symmetry group is doubly transitive, then $C>0$ implies $\lambda_{\min} > 0$.
Otherwise, $q$ is prime and $\lambda_{\min} >0$ by assumption. The desired convergence is established by the following result, which is obtained by combining the rate difference property of GRM codes in Lemma~\ref{lem:grm_rate_diff} with the generic bound in Lemma~\ref{lem:rate_cover}.

\begin{lemma}%
\label{lem:dexit_GRM}
For a GRM code RM$_q (r,m)$ with $m\geq q^2$ on a channel $W$ whose overlap matrix has minimal eigenvalue $\lambda_\mathrm{min}>0$, we find that
\begin{align*}
\dexit_\mathrm{avg}
& \leq \frac{2 \ln q}{\lambda^2_\mathrm{min} } \left( \frac{7+3\log_q m }{\sqrt{m}} \right) = O \left(\frac{\ln m}{\sqrt{m}} \right).
\end{align*}
\end{lemma}

\vspace{0.5mm}

\ifarxiv

\subsection{Proof of Corollary~\ref{cor:main}}
For this result, we use the idea of group symmetrization in Section~\ref{sec:group_sym}.
Applying this operation to $W$ with the affine group $A_{\q}$ over $\mathbb{F}_{\q}$ forces the symmetrized channel to satisfy the condition \emph{(i)} in Theorem~\ref{thm:main}.
Then, applying Theorem~\ref{thm:main} proves the corollary.
We note that this channel symmetrization operation is the same as multiplying the codeword elementwise by a uniform random $\mathbb{F}_{\q}^*$ vector and then adding elementwise a uniform random $\mathbb{F}_q$ vector.
These vectors are shared with the receiver in advance so that the process can be inverted during decoding. \qed

\fi

\bibliographystyle{ieeetr}

\ifarxiv

\appendices

\section{Generalized Reed--Muller Codes}
\label{sec:grm_codes}

In this section, we describe GRM codes in more detail.
For a vector $\bm{v}=(v_0,\ldots,v_{m-1})$ of $m$ indeterminates, let $\bm{d}\in \mathbb{N}_0^m$ define the monomial 
\[\bm{v}^{\bm{d}} \coloneqq \prod_{i=0}^{m-1} v_i^{d_i}.\]
Let $\mathbb{F}_q [v_0,\ldots,v_{m-1}]$ be the vector space (over $\mathbb{F}_q$) of $m$-variate polynomials with coefficients in $\mathbb{F}_q$.
This space is spanned by the monomials $\bm{v}^{\bm{d}}$ with $\bm{d}\in \mathbb{N}_0^m$ and the degree of a monomial is given by \[\deg(\bm{v}^{\bm{d}}) \coloneqq \sum_{i=0}^{m-1} d_i . \]

Codewords will be formed by evaluating elements of $F_q [v_0,\ldots,v_{m-1}]$ at all points $\bm{v} \in \mathbb{F}_q^m$. 
Since $\alpha^{q} = \alpha$ for all $\alpha\in \mathbb{F}_q$, we restrict our attention to exponent vectors $\bm{d}\in [q]^m$ whose elements are at most $q-1$.
Thus, for RM$_q (r,m)$, the set of relevant exponent vectors are given by
\[ \cM_{q,r,m} \coloneqq \left\{  \bm{d} \in [q]^m \;  \middle| \;  \sum_{i=0}^{m-1} d_i \leq r \right\}. \]
To continue, we need an ordering for the elements of $\mathbb{F}_q^m$.
For now, we assume only that $\tau_m \colon [q^m] \to \mathbb{F}_q^m$ is a bijective function (i.e., that enumerates $\mathbb{F}_q^m$).
Then, the evaluation of a monomial is defined by %
\[ \mathrm{ev} (\bm{v}^{\bm{d}} ) \coloneqq  (\tau(0)^{\bm{d}},\tau(1)^{\bm{d}},\ldots,\tau(q^m-1)^{\bm{d}}). \]
Using this, the code RM$_q (r,m) \subseteq \mathbb{F}_q^N$ is the subspace spanned by the vectors $\mathrm{ev}(\bm{v}^{\bm{d}})$ for all $\bm{d}\in \cM_{q,r,m}$.

To define a specific ordering for $\mathbb{F}_q^m$, we choose a fixed primitive element $\beta \in \mathbb{F}_q$ and represent $\mathbb{F}_q$ as the ordered set $(0,1,\beta,\beta^2,\ldots,\beta^{q-2})$.
Then, we can define $\tau_1 \colon \mathbb{F}_q \to [q]$ to extract the rank of an element via
\[v \mapsto \begin{cases}
0 & \text{ if $v=0$} \\
i+1 & \text{ if $v=\beta^i$}. \\
\end{cases}
\]
Using this, a reverse lexicographical ordering of vectors is given by the function $\tau_m \colon \mathbb{F}_q^m \to [q^m]$ satisfying
\[ \tau_m (\bm{v} ) = \sum_{i=0}^{m-1} \tau_1 (v_i) \, q^{i}. \]
When $m$ is clear from the context, we will drop the subscript.
We note that the order $\tau_1$ is defined explicitly for sake of concreteness but the only necessary property is that $\tau_1 (0)=0$.

A key property of polynomial evaluation codes is that, for an invertible matrix $Q\in\mathbb{F}_{q}^{m\times m}$ and a vector $\bm{b} \in\mathbb{F}_
{q}^{m}$, the degree of a polynomial is preserved by the affine change of variables $\bm{v} \mapsto \pi_{Q,\bm{b}}(\bm{v})$ where  $\pi_{Q,\bm{b}}\colon\mathbb{F}_q^{m}\to \mathbb{F}_q^{m}$ is defined by
\[
[\pi_{Q,\bm{b}} (\bm{v})]_{i}
= \sum_{j=1}^{m}Q_{i,j}v_{j}+b_{i}.
\]
Thus, the set of polynomials formed by linear combinations of monomials with exponent vectors in $\cM_{q,r,m}$ is mapped to itself by this change of variables and, thus, the permutation $\tau (\pi_{Q,\bm{b}}(\tau^{-1} (i)))$ defines an automorphism of RM$_q (r,m)$ in terms of symbol indices~\cite{Kasami-it68}.

\subsection{Rates of Generalized Reed--Muller Codes.}

Let $\bm{D} \in \{0,1,\ldots,q-1\}^m$ be a random variable distributed uniformly on the set $\{0,1,\ldots,q-1\}^m$.
Then, the degree of a uniform random monomial (with exponents at most $q-1$) 
 satisfies $\deg(\bm{v}^{\bm{D}}) = \sum_{i=0}^{m-1} D_i$ is equal the sum of $m$ i.i.d.\ random variables $D_0,\ldots,D_{m-1}$ which are distributed uniformly on the set $\{0,1,\ldots,q-1\}$.
The rate of the GRM code RM$_q (r,m)$ equals the fraction of monomials with degree at most $r$ and this is given by the cumulative distribution function (cdf) of this sum
\[ R_q (r,m) \coloneqq \frac{1}{q^m} | \cM_{q,r,m} | = \Pr \left[\deg(\bm{v}^{\bm{D}}) \leq r \right]. \]

Here we place the proofs of Lemma~\ref{lem:grm_rate_diff} and Lemma~\ref{lem:grm_punct}

\begin{proof}[Proof of Lemma~\ref{lem:grm_rate_diff}]
To understand the rate of GRM codes, we will apply the Berry-Esseen central limit theorem to analyze the number of monomials with degree at most $r$.
To do this, we first list the relevant moments:  $ \mu \coloneqq \ex{D_0} = \frac{q-1}{2}$,
\begin{align*}
\sigma^2 &\coloneqq \ex{\left(D_0 - \frac{q-1}{2} \right)^2} = \frac{q^2 -1 }{12}, \\
\rho &\coloneqq \ex{\left| D_0 - \frac{q-1}{2} \right|^3} = \begin{cases} \frac{1}{32} q \left(q^2-2\right) & \text{ if $q$ is even,} \\
\frac{1}{32 q} \left(q^2-1\right)^2 & \text{ if $q$ is odd.}
\end{cases}
\end{align*}
Let $\Phi(\alpha) = (2\pi)^{-1/2} \int_{-\infty}^\alpha \exp( - z^2/2) \, dz$ be the cdf of a standard Gaussian random variable.
The Berry-Esseen central limit theorem (with constant $\frac{1}{2}$ justified by~\cite{Shevtsova-inf2013}) states that
\begin{align}
\sup_{x\in\mathbb{R}} \bigg| \mathbb{P} & \left[\deg(\bm{v}^{\bm{D}}) \leq r \right] - \Phi\left(\frac{r-m\mu}{\sigma \sqrt{m}} \right) \bigg| \\ &\leq \frac{\rho}{2\sigma^3 \sqrt{m}} \\
&= \begin{dcases}
\frac{\sqrt{27} q \left(q^2-2\right)}{8 \left(q^2-1\right)^{3/2} \sqrt{m}} & \text{ if $q$ is even} \\
\frac{\sqrt{27} \sqrt{q^2-1}}{8 q \sqrt{m}} & \text{ if $q$ is odd} \end{dcases}
\\&\leq \frac{1}{\sqrt{m}}, \label{eq:grm_berry_esseen}
\end{align}
where the last step holds because both expressions are increasing in $q$ and thus, as $q\to\infty$, they are upper bounded by their common limit $\sqrt{27}/(8\sqrt{m}) \approx 0.649/\sqrt{m}$.
If we consider a sequence of GRM codes where $n$-th code is RM$_q (r_{n},m_{n})$ with $m_{n} \to \infty$ and $r_n = m\mu+\alpha \sigma \sqrt{m} + o(\sqrt{m_n})$, then this implies that
\[ R_q (r_n,m_n) \to \Phi(\alpha). \]

To simplify notation, we now prove a modified version of the stated lemma formed by mapping $m \mapsto m+k$.
Let $\alpha(r,m)=\frac{r-m\mu}{\sigma \sqrt{m}}$. 
For $k=0$, the statement is trivial.
For $k\geq 1$, we start with~\eqref{eq:grm_berry_esseen} to see that
\begin{align*}
R_q &(r,m) - R_q(r,m+k) \\
&\leq \Phi\left(\alpha(r,m) \right) - \Phi\left(\alpha(r,m+k) \right) + \frac{1}{\sqrt{m}} + \frac{1}{\sqrt{m+k}}\\
&\leq \frac{2}{\sqrt{m}} + \int_{m+k}^m \left( \Phi' \left(\alpha(r,z) \right) \frac{d}{dz} \alpha(r,z) \right) dz \\
&\overset{(a)}{\leq} \frac{2}{\sqrt{m}} + \frac{1}{\sqrt{2\pi}} \int_{m}^{m+k}  \frac{z \mu +r}{2 z^{3/2} \sigma} dz \\
&\overset{(b)}{\leq} \frac{2}{\sqrt{m}} + \frac{1}{\sqrt{2\pi}} \int_{m}^{m+k}  \frac{z(q-1)\sqrt{3} + m(q-1)\sqrt{12}}{2 z^{3/2} \sqrt{q^2-1}} dz \\
&\overset{(c)}{\leq} \frac{2}{\sqrt{m}} + \frac{1}{\sqrt{2\pi}} \int_{m}^{m+k}  \left(\frac{\sqrt{12}}{4 \sqrt{z}}+\frac{m\sqrt{12}}{2 z^{3/2}}\right) dz \\
&\overset{(d)}{\leq} \frac{2}{\sqrt{m}} + \frac{3\sqrt{3} k}{2\sqrt{2\pi}\sqrt{m}} \leq \frac{4k}{\sqrt{m}},
\end{align*}
where $(a)$ holds because $\Phi'(x) \leq 1/\sqrt{2\pi}$ for all $x\in\mathbb{R}$, $(b)$ follows from $r\leq m(q-1)$, $(c)$ is given by $(q-1)/\sqrt{q^2-1}=\sqrt{q-1}/\sqrt{q+1}\leq 1$, $(d)$ holds because $z\geq m$ for $z\in [m,m+k]$, and the last step follows because we have $2+ck\leq 4k$ for $k\geq 1$ and $c=3\sqrt{3}/(2\sqrt{2\pi})\approx 1.04$.
\end{proof}

\begin{proof}[Proof of Lemma~\ref{lem:grm_punct}]
To see this, one can split the monomials into two groups.
Let the first set of monomials be those that only contain the variables $v_0,\ldots,v_{m-k-1}$ (i.e., with exponent vectors in $\cM_{q,r,m}$ where the last $k$ elements are 0).
Notice that this equals the set of monomials given by $\bm{v}^{\bm{j}}$ for  $\bm{j}\in \cM_{q,r,m-k}$.
That means the second set, which  contains all the rest, is given by $\cM' = \cM_{q,r,m} \setminus \cM_{q,r,m-k}$.
Let $V=\mathbb{F}_q^{m-k}\times\left\{ 0\right\} ^{k}$ and observe that $I = \tau(V) = [q^{m-k}]$. 
The key observation is that the monomials in $\cM'$ all evaluate to 0 on the set $V$ because all points in $V$ have $v_{m-k}=\cdots=v_{m-1}=0$.
Thus, for $\bm{c} \in \cC$ and $i\in I$, only the monomials in $\cM_{q,r,m-k}$ contribute to the value of $c_i$.
This implies that the codewords in $\cC_I$ are formed by evaluating the polynomials formed by linear combinations of monomials with exponent vectors in $\cM_{q,r,m-k}$ (i.e., that only depend on $v_0,\ldots,v_{m-k-1}$ and have total degree at most $r$).
Moreover, this notation orders the vector $c_I$ so that the evaluation at $\bm{v} \in V$ appears before the evaluation at $\bm{v}' \in V$ iff $\tau(\bm{v}) < \tau(\bm{v}')$. 
This is a reverse lexicographic ordering on $(v_0,\ldots,v_{m-k-1})$ and, hence, $\cC_I$ is precisely equal to $\mathrm{RM}_q(r,m-k)$.
Another important point is that exactly $q^{|\cM'|}=|\cC|/|\cC_I|$ codewords in $\cC$ are mapped to each codeword in $\cC_I$.
This holds because, if the information symbols associated with $\cM_{q,r,m-k}$ are fixed, then the punctured codeword $c_I$ is fixed.
But, by choosing the information symbols associated with the monomials in $\cM'$, one can generate $q^{|\cM'|}$ different codewords in $\cC$ that have the same $c_I$.
\end{proof}

\section{Properties of the Overlap Matrix}\label{sec:overlap_properties}

One natural interpretation of the overlap matrix (introduced in Section~\ref{sec:overlap}) is that it is the probability transition matrix for the degraded channel obtained by sampling according to the canonical map. Specifically, consider a Markov chain $X- Y - X'$ where $Y$ is an observation of $X$ through channel $W$ and  $X'$ is drawn according to the canonical map $\phi(Y)$. For all $x,x' \in \cX$, we have
\begin{align}
\quad\pr{X' = x' \mid X = x} %
&= \int \phi_{x'}(y)  W(dy \mid x) \\
& = \int  \phi_{x'}(y)  \phi_{x}(y)  \bar{W}(dy) \\
& \eqqcolon Q_{x,x'}.  \label{eq:Wsampled}
\end{align}

It is interesting to note that the channel defined by this sampling procedure i.e., the channel with input $X$ and output $X'$, has overlap matrix $Q^2$. Thus, the discrepancy $\delta = \frac{1}{q} \gtr(Q) - \frac{1}{q} \gtr( Q^2) $ can be interpreted as the degradation in the overlap matrix due to sampling the output.

\subsection{Principle Inertia Components} 
Let $(X,Y)$ be an input-output pair through a $q$-ary channel $W$ where $X$ is uniformly distributed.  From the definition of canonical map and the overlap matrix, one finds that  $(1, \sqrt{\lambda_1},  \dots ,  \sqrt{\lambda_{q-1}})$ are the singular values of the conditional expectation operator  $T_Y \colon L_2(\cX, P_X) \to L_2(\cY, P_Y)$ defined by 
\[
(T_Y f)(y) = \ex{ f(X) \mid  Y =y}.
\]
By \cite[Theorem~1]{calmon:2017} (see also \cite[Section~3]{witsenhausen:1975}) it follows that, $\lambda_{1}, \dots, \lambda_{q-1}$ are the PICs for the channel $W$ with uniform input distribution. The square root of the largest PIC is also known as the Hirschfeld-Gebelein-R\'{e}nyi maximal correlation between the channel input and output.

Let $P_{XY}$ denote the joint probability measure of the input-output pair and let $P_X \otimes P_Y$ be the product measure with the same marginals. The mutual information $I(X;Y)$ and the $\chi^2$-information $\chi^2(X;Y)$ are defined by 
\begin{align}
I(X;Y) &\coloneqq D( P_{X,Y} \, \| \, P_{X} \otimes P_Y)\\
\chi^2(X;Y) &\coloneqq \chi^2( P_{X,Y} \, \| \, P_{X} \otimes P_Y),
\end{align}
respectively, where, for probability  measures $P \ll Q$,  $D(P \, \|\,  Q) \coloneqq \int (\log dP/dQ) dP$ is the Kullback-Leibler divergence and $\chi^2(P \, \| \, Q) \coloneqq \int  (dP/dQ)^2dP  -1$ is the $\chi^2$-divergence. 
It is well known (see \cite[Section~I-C]{calmon:2017}) that
\begin{align}
    \chi^2(X;Y) = \sum_{x=1}^{q-1} \lambda_x = \gtr(Q) - 1. 
\end{align}
In view of the general inequality $D(P \, \| \, Q)\le \log(1 + \chi^2( P\, \| \,Q)$ it follows that $I(X; Y) \le \log \gtr(Q) $.

\subsection{Overlap for Symmetric Channels}

By construction, the overlap matrix is invariant to simultaneous permutation of its columns and rows by permutations in the symmetry group of the channel, i.e., 
\begin{align}
 \forall \sigma \in G, \forall x,x' \in \cX, \quad    Q_{x,x'} =  \kexit_{\sigma x, \sigma x'}. \label{eq:Q_sym} 
\end{align} 

\begin{lemma}\label{lem:Qsym}
Let $Q$ be the overlap matrix of a $q$-ary channel with symmetry group $G$. 
\begin{enumerate}[(i)]
    \item If $G$ is transitive, then $Q$ has constant diagonal entries and each column (row) is a permutation of the first column (row). Furthermore, the maximum value in each column (row) occurs on the diagonal. 

    \item If $G$ is doubly transitive, then $\kexit$  has constant diagonal entries and constant off-diagonal entries.
\end{enumerate}
\end{lemma}
\begin{proof}
First we consider part (i). The transitivity of $G$ implies that for each $x \in\cX$, there exists a permutation $\sigma \in G$ mapping $x$ to $0$. In view of \eqref{eq:Q_sym}, it follows that $Q_{u,x} = Q_{\sigma u, \sigma x}  = Q_{\sigma u,0}$ for all $u \in \cX$, 
and so the $x$-th column is a permutation of the $0$-th column. This relation also implies that  $Q_{x,x}= Q_{0,0}$ for all $x$ and so the diagonal entries are identical.  To see that  the maximum value in each column occurs on the diagonal, we use the fact that $Q$ is symmetric and positive semidefinite and thus $ Q_{x,x}  Q_{0,0} - Q_{x,0}^2 \ge 0  $  for all $x \in\cX$. 
In conjunction with the fact that $Q_{x,x} = Q_{0,0}$ we conclude that the maximum value occurs on the diagonal.
Part (ii) follows directly from~\eqref{eq:Q_sym}.
\end{proof}

For a discrete channel defined by an $|\cX|\times |\cY|$ row-stochastic transition matrix $W$ with no all-zero columns, the overlap matrix is given by $Q=W \diag(\one^T W)^{-1} W^T$.

\begin{remark}
For a non-symmetric channel it is possible that the maximum value in a column (row) of the overlap matrix does not occur on the diagonal. For example, we can compute the overlap matrix $Q$ for the case where $\cX = \{0,1,2,3\}$, $\cY = \{a,b\}$  and the channel transition matrix is $W$:
\begin{align}
 W = \begin{pmatrix} 1 &  0\\[1pt]
\frac{1}{2} & \frac{1}{2} \\[1pt]
0 & 1\\[1pt]
0 & 1
\end{pmatrix}
\;\; \implies \;\;
Q  = \begin{pmatrix}
 \frac{2}{3} & \frac{1}{3} & 0 & 0 \\[1pt]
 \frac{1}{3} & \frac{4}{15} & \frac{1}{5} & \frac{1}{5} \\[2pt]
 0 & \frac{1}{5} & \frac{2}{5} & \frac{2}{5} \\[1pt]
 0 & \frac{1}{5} & \frac{2}{5} & \frac{2}{5} \\[1pt]
\end{pmatrix}.
\end{align}
\end{remark}

\section{Proofs of Results in Section~\ref{sec:sym_channels}}

\begin{proof}[Proof of Lemma~\ref{lem:sylow_group}]
Let $q=p^m$ for some prime $p$ and integer $m$.
Since $G$ is transitive (by channel symmetry), one can show that $q$ divides $|G|$ by observing that the $G$-orbit of any $x \in \cX$ has $q$ elements and then applying the orbit-stabilizer theorem.
Then, by Sylow's First Theorem, there is a Sylow $p$-subgroup $H\subseteq G$ with $q$ elements. Also, if a group acts transitively on a set, then so does any Sylow-$p$ subgroup~\cite[p.~121]{milneGT}. Thus, $H$ is a transitive subgroup with $q$ elements and $W$ is group symmetric.
If $q=p$, then the group structure of $H$ is unique and $H$ is isomorphic to the the cyclic group with $q$ elements.
If $q=p^2$, then there are two possibilities for the group structure of $H$; it is either the cyclic group with $q=p^2$ elements or the direct product of two cyclic group with $q=p$ elements~\cite[Prop.~7.3.3]{Artin-2011}.
In both of these cases, $H$ is abelian.
In general, however, $H$ may be non-abelian.
\end{proof}

\begin{proof}[Proof of Lemma~\ref{lem:SERtoOverlap}]
Let $X'$ be sampled according to the canonical map $\phi(Y)$ such that $X - Y - X'$ is a Markov chain. From the Markov structure and the definition of the SER we have
\begin{align}
   \ser(X \mid Y) & \le \ser(X \mid X') \le  1-   \pr{X' = X}. 
\end{align}
From \eqref{eq:Wsampled}, it follows that
\begin{align}
 \pr{ X' =  X}
& = \sum_{x \in \cX}p_x \pr{ X' =x \mid X = x}  = \sum_{x \in \cX} p_x Q_{x,x},
\end{align}
and so the proof is complete. 
\end{proof}

\begin{proof}[Proof of Lemma~\ref{lem:strong_concavity}]
Let $X'$ be sampled according to the canonical map $\phi(Y)$ such that $S - T - X - Y - X'$ is a Markov chain. By \eqref{eq:Wsampled}, the conditional pmf of $X'$ given $X = x$ is the $x$-th column of the overlap matrix and thus 
\begin{align}
\pr{X' = x' \mid X} = e_x^\top Q \be_X.
\end{align}
For $U \in \{S,T\}$, the conditional probability of $X'$ given $U$ is then obtained by taking the conditional expectation: 
\begin{align}
\pr{X' = x' \mid U} = e_x^\top Q  \ex{ \be_X \mid U}. \label{eq:X'cond}
\end{align}

Meanwhile, for $a,b \in \Delta_q$, we can use Pinsker's inequality followed by the inequality $\|\cdot\|_{\ell_1} \le \|\cdot\|_{\ell_2}$ to write
\begin{align}
    D( a \, \| \, b)  \ge \frac{1}{2 \ln q} \|a - b\|^2_1  \ge \frac{1}{2 \ln q} \|a - b\|^2_2, \label{eq:Pinsker}
\end{align}
where $D(\cdot \, \| \, \cdot)$ denotes the Kullback-Leibler divergence defines by probability vectors. 

Using these properties, we can write
\begin{align*}
\MoveEqLeft I(X ; Y \mid S) - I(X ; Y \mid T)\\
& =  I(T;Y \mid S) \\
&\overset{(a)}{\ge}   I(T; X' \mid S)  \\
&\overset{(b)}{=} H(X' \mid S) - H(X' \mid T) \\
&\overset{(c)}{=} \ex{\sum_{\;x' \in \cX} \pr{X' = x' \mid T} \log \frac{\pr{X' = x' \mid T}}{\pr{X' = x' \mid S}}} \\
&\overset{(d)}{=} \ex{ D( \kexit \ex{\be_X \mid T} \, \| \,  \kexit  \ex{ \be_X \mid S} ) )} \\
& \overset{(e)}{\ge} \frac{1}{2 \ln q} \ex{  \| \kexit  ( \ex{\be_X \mid T}  -  \ex{ \be_X \mid S} )) \|^2}\\
& \overset{(f)}{\ge} \frac{\lambda^2_{q-1}}{2 \ln q} \ex{  \|  \ex{\be_X \mid T}  -  \ex{ \be_X \mid S} )\|^2},
\end{align*}
where $(a)$ is the data processing inequality, $(b)$ follows from the Markov chain condition, $(c)$ is given by the definition of entropy, $(d)$ hold because of \eqref{eq:X'cond} and the definition of KL divergence, $(e)$ follows from \eqref{eq:Pinsker}, and $(f)$ holds because $Q$ is symmetric with minimum eigenvalue $\lambda_{q-1}$.
\end{proof}

\begin{proof}[Proof of Lemma~\ref{lem:Bconstraints}]
Define the set $B \coloneqq \{x \in \cX : \lambda_x \ge 1/2\}$. Starting with \eqref{eq:dexit_alt}, we can write   
\begin{align}
\dexit & = \frac{1}{q}  \sum_{x \in B}  \lambda_x ( 1 - \lambda_x) + \frac{1}{q}\sum_{x \notin B}  \lambda_x ( 1- \lambda_x) \notag\\
& \ge  \frac{1}{2q} \sum_{x \in B}   |  \lambda_x - 1|  + \frac{1}{2q} \sum_{x \notin B}    |\lambda_x |   = \frac{1}{2q} \| \lambda - \one_B \|_1. \notag
\end{align}
By the reverse triangle inequality, 
\begin{align}
2 \q \dexit & \ge \left| \| \lambda \|_1 - \| \one_B\|_1 \right| = 
\left| \gtr(\kexit) - |B| \right|. \label{eq:TrQ2B}
\end{align}

\paragraph*{Case (i)}
If the symmetry group is doubly transitive, then part (ii) of Lemma~\ref{lem:Qsym} implies that the eigenvalues of the overlap matrix with index greater than 0 are identical, i.e., $\lambda_1 = \lambda_2 = \dots = \lambda_{q-1}$.  In conjunction with the fact that $\lambda_0 =1$ (because the overlap matrix is doubly stochastic), we see that the only possibilities are $|B| \in \{1,q\}$, and thus \eqref{eq:trQtob} follows from \eqref{eq:TrQ2B}.  

\paragraph*{Case (ii)}
Next, we consider the case where the symmetry group is  transitive but not necessarily doubly transitive.  By part (i) of  Lemma~\ref{lem:Qsym}, every column of the overlap matrix is a permutation of the the first column, which we will denote by $\mu =  (\mu_0, \dots, \mu_{\q-1})$. Furthermore, the largest entry in each row occurs on the diagonal and is equal to $\mu_0$. Starting with \eqref{eq:dexit_alt} and then noting that $\gtr(\kexit) = q \mu_0$ and $\gtr(\kexit^2) = \q \|\mu\|^2$, leads to 
\begin{align*}
\dexit = \mu_0  - \| \mu\|^2 = \sum_{x \in \cX} \mu_x (\mu_0 - \mu_x).
\end{align*}
To proceed define the set $A \coloneqq \{ x \in \cX : \mu_x \ge \mu_0/2\}$ and observe that since  $\mu_0$ is largest entry in $\mu$, we have 
\begin{align}
\dexit & = \sum_{x \in A}  \mu_x ( \mu_0 - \mu_x) + \sum_{x \notin A}  \mu_x ( \mu_0 - \mu_x) \notag\\
& \ge \frac{\mu_0}{2}   \sum_{x \in A}  |  \mu_0   - \mu_x|  +  \frac{\mu_0}{2}  \sum_{x \notin A}  |  \mu_x | \notag\\
&= \frac{\mu_0}{2} \| \mu - \mu_0 \one_A\|_1. 
\end{align}
By the reverse triangle inequality, 
\begin{align}
\frac{2}{\mu_0} \dexit & \ge \left| \|\mu\|_1 - \| \mu_0 \one_A\|_1 \right|  =  \left| 1-  \frac{1}{\q} \gtr(\kexit) |A| \right|.\label{eq:TrQ2A}
\end{align}

Combining \eqref{eq:TrQ2B} and \eqref{eq:TrQ2A} with $1/\mu_0 \le q$ leads to 
\begin{align*}
\left |\q -  |A| |B|  \right|  \le  \frac{ 2  \q \dexit }{\mu_0}  + 2 \q |A| \dexit \le 4 \q^2 \dexit.
\end{align*}
From this inequality, we observe that if  $4 q^2 \dexit < 1$, then $|A|$ and $|B|$ must be integer factors of $q$. If $q$ is prime then the only possibilities are $|B| \in \{1,q\}$, and thus \eqref{eq:trQtob} follows from~\eqref{eq:TrQ2B}. 
\end{proof}

\section{Proofs of Results in Section~\ref{sec:codes}}
\label{app:group_codes}

\begin{proof}[Proof of Lemma~\ref{lem:coset_chan_sym}]
To simplify notation we set $n = N-1$ and  assume,  without loss of generality,  that $W$ is a standard channel with output space $(\simp{q}, \cB)$. Let  $\{\sigma_x : x \in \cX\}$ be the  group of permutations on $\cX$ defined by $\sigma_{x} x' = x \gx x'$. Similarly, let $\{ \pi_{\bx} : \bx \in \cX^n\}$ be the group of permutations on $\cX^n$ defined by  $\pi_{\bx} \bx' = \bx \gx \bx'$, and let the operation of $\pi_{\bx}$ on  $\Delta_q^n$ be given by 
\begin{align}
    \pi_{\bx} (y_0, \dots, y_{n-1})  &\coloneqq (\sigma_{x_0} y_0, \dots, \sigma_{x_{n-1}} y_{n-1}) 
\end{align}

We use $W^n$ to denote the product channel extension of the channel $W$ to a conditional product measure on $(\simp{\q}^{n},  \cE)$,
satisfying 
\[ 
W^{n} \left( \bigtimes_{i=0}^{n-1} B_i \, \middle| \, \bm{x} \right)
= \prod_{i=0}^{n-1} W(B_i \mid x_i), 
\]
for all $\bx \in \cX^n$ and  $B_0,\ldots,B_{n-1}\in \cB$. Using this notation, the coset channel can be written explicitly as 
\[ 
\cch (E \mid u ) \coloneqq \frac{|\cX|}{|\cC|} \sum_{\bx \in \cC : x_0 = u } W^{n} (E \mid x_{\sim 0}), 
\]
for all $u \in \cX$ and $E \in \cE$. Here, the condition that, for all $u \in \cX$, there exists  $\bc \in \cC$ with $c_0 = u$ ensures that this channel is well-defined on the input  alphabet $\cX$. 

The condition that the symmetry group of $W$ contains the permutation representation of $(\cX, \gx)$ implies that the symmetry group of  $W^{n}$ contains the permutation representation  of the product group. In particular, we can write
\begin{align}
W^{n} (E \mid \bx)
&= W^{ n} (\pi_{\bx'} E  \mid   \pi_{\bx'} \bx  ),
\label{eq:Wn_sym}
\end{align}
for all $\bx, \bx' \in \cX^n$ and $E \in \cE$. 

To see that  the coset channel inherits the symmetry of the channel symmetry group,  observe that for all $u \in \cX$ and $\bc \in \cC$, 
\begin{align}
\cch  (E \mid u) 
& = \frac{|\cX|}{|\cC|} \sum_{\bx \in \cC \,:\, x_0 = u } W^{n} (E \mid x_{\sim 0}),\\
&= \frac{|\cX|}{|\cC|} \sum_{\bx \in \cC \,:\, x_0 = u } W^{n} (\pi_{c_{\sim 0}}  E \mid \pi_{c_{\sim 0}} x_{\sim 0} ),\\
&= \frac{|\cX|}{|\cC|} \sum_{\bx \in \cC \,:\, x_0 =  \sigma_{c_0} u } W^{n} (\pi_{c_{\sim 0}}  E  \mid x_{\sim 0}),\\
&= \cch  (\pi_{c_{\sim 0}}  E  \mid \sigma_{c_0} u ), \label{eq:V_sym}
\end{align}
where the third equality uses the fact that group codes are closed under addition (i.e., under permutations  $\{\pi_{\bc} : \bc \in \cC\}$). 

Using the compact notation $V_u(\cdot) = V(\cdot \mid u)$, this symmetry condition can be expressed as
\begin{align}
V_{\sigma_{c_0} u} = \pi_{c_{\sim 0}} V_{u}  , \quad  \forall u \in \cX, \forall \bc \in \cC.
\end{align}
Recalling that the canonical map  $\psi \colon \Delta_q^n \to \Delta_q$ is defined by  $\psi \coloneqq ( d V_u  /d \bar{V} \mid u \in \cX)$ and then noting that the reference measure $\bar{V}$ is invariant to transformations by $\pi_{c_{\sim 0}}$,  we see that $\psi$ has the same symmetries as the channel with the roles of the input and the output interchanged: 
\begin{align}
\sigma_{c_0} \psi(\cdot ) =  \psi( \pi_{c_{\sim 0}} (\cdot) ) , \quad   \forall \bc \in \cC. 
\end{align}
In particular, for $B \in \cB$ and $\bc \in \cC$, this implies
\begin{align}
\psi^{-1} B & = \{ y_{\sim 0} \in \Delta^n_q \mid  \psi(y_{\sim 0}) \in B \}\\
& = \{ y_{\sim 0} \in \Delta^n_q \mid  \psi( \pi_{c_{\sim 0}} y_{\sim 0}) \in  \sigma_{c_0} B \}\\
& = \pi_{c_{\sim 0}}^{-1} \psi^{-1} \sigma_{c_0} B. \label{eq:psi_sym}
\end{align}

Evaluating \eqref{eq:V_sym} with $E=\psi^{-1} B$ and then applying \eqref{eq:psi_sym}, leads to
\[
V( \psi^{-1} B \mid u)  = V( \psi^{-1} \sigma_{c_0}  B \mid \sigma_{c_0} u)
\]
and thus the symmetry group of the coset channel contains the permutation representation of the group $(\cX, \gx)$.

Lastly, we show that the distribution of the canonical map is conditionally independent of the coset representative given $X_0$. The group structure of the code implies that for any  $\bx, \bx' \in \cC$ there exists a codeword $\bc \in \cC$ mapping $\bx$ to $\bx'$. In particular, if $\bx$ and $\bx'$ are in the same coset (i.e., $x_0 = x_0'$) then $c_0 =0$ is the identity element. Evaluating \eqref{eq:Wn_sym} with $E=\psi^{-1} B$ and then applying \eqref{eq:psi_sym} leads to 
\begin{align}
\MoveEqLeft \pr{ \psi(Y_{\sim 0}) \in B \mid \bX = \bx}\\
& =  W^n( E \mid x_{\sim 0}) \\
& =  W^n( \pi_{c_{\sim 0}} E \mid \pi_{c_{\sim 0}} x_{\sim 0}) \\
& =  W^n( \psi^{-1} \sigma_{c_0} B   \mid \pi_{c_{\sim 0}} x_{\sim 0}) \\
 & = W^n( \psi^{-1}  B \mid   x'_{\sim 0}) \\
& = \pr{ \psi(Y_{\sim 0}) \in B \mid \bX = \bx'}. 
\end{align}
Therefore, the distribution of $\psi(Y_{\sim 0})$ depends only on $X_0$. This completes the proof. 
\end{proof}

\begin{proof}[Proof of Lemma~\ref{lem:HF_matched}]
We use the same notation as in the proof of Lemma~\ref{lem:coset_chan_sym}. The fact that $F'$ is a subgroup of $G$  means that the product channel $W^n$ satisfies
\begin{align}
W^n(E \mid \bx) = W^n( \sigma E \mid \sigma \bx)  
\end{align}
for all $\bx \in \cX^n$,  $E \in \cE$, and $\sigma \in F'$.  Following the same steps leading to \eqref{eq:V_sym}, verifies that $
V(\sigma E \mid \sigma u) = V(E \mid u)$,
for all $u \in \cX$,  and thus the symmetry group of the coset channel contains $F'$.

Finally, because groups are closed, the symmetry group of $V$ must contain all elements generated by its subsets.
Thus, the symmetry group of $V$ must contain $\langle H,F' \rangle$.
\end{proof}

\begin{proof}[Proof of Lemma~\ref{lem:linear_code_affine}]
\label{proof:linear_code_affine}
As $\cC \subseteq \mathbb{F}_q^N$ is a linear code, it is a subspace of $\mathbb{F}_q^N$ and $a \bc + \bc' \in \cC$ for all $\bc, \bc' \in \cC$ and $a\in \mathbb{F}_q$.
Since it contains the all ones codeword $\one \in \cC$, it follows that $\bc + b \one \in \cC$ for all $b\in \mathbb{F}_q$.
Putting these together, we find that $a \bc + b \one \in \cC$ for all $\bc \in \cC$ and $a,b\in \mathbb{F}_q$.
From this, we see that the set $F$ of homogeneous alphabet relabelings for $\cC$ must contain the affine group $A_q$, which is doubly transitive. 
Applying Lemma~\ref{lem:HF_matched} completes the proof.
\end{proof}

\section{Proofs of Results in Section~\ref{sec:main}}
\label{sec:proofs}

\begin{proof}[Proof of Lemma~\ref{lem:weak_bound}] 
The assumptions on the input distribution and the channel 
combined with the area theorem  for the mutual information in Lemma~\ref{lem:area} (evaluated with $S = [N]$) provide a lower bound on the code rate:
 \begin{align}
R &= \frac{H(\bX)}{N} \ge 
\frac{I(\bX ; \bY)}{N} \\
& =     \int_0^1 I( X_0 ; Y_0 \mid  Y_{\sim 0}(s) ) \, ds.\label{eq:weakRLB}  
 \end{align}

Meanwhile, for all $0 \le s \le 1$, the channel capacity satisfies the upper bound: 
\begin{align}
C  & = I(X_0 ; Y_0) \\
& = I(X_0, Y_{\sim 0}(t) ; Y_0 ) \\
& = I(X_0 ; Y_0 \mid  Y_{\sim 0}(s)  ) + I(Y_0; Y_{\sim 0}(s)  )\\
& \le I(X_0 ; Y_0 \mid  Y_{\sim 0}(s)  ) + I(X_0; Y_{\sim 0}(s)  ).\label{eq:weakCexp}  
\end{align}
The first equality holds because the capacity of a symmetric channel is attained by the uniform  distribution on its input~\cite[Theorem~21]{Polyanskiy-it13}. The remaining steps follow from the chain rule for mutual information, the fact that $Y_0 - X_0 - Y_{\sim 0}(t)$ is a Markov chain, and the data processing inequality. 

Rearranging \eqref{eq:weakCexp} and  observing that,  for $0 \le t \le s$,  the channel from $X_0$ to $Y_{\sim 0}(s)$ is degraded with respect to the channel from $X_0$ to $Y_{\sim 0}(t)$ leads to
\begin{align}
I(X_0 ; Y_0 \mid  Y_{\sim 0}(s)  ) \ge \left(C  - I(Y_0; Y_{\sim 0}(t)) \right) \one_{\{s \ge t\}}.\label{eq:weakCexp2}
\end{align}
Using this inequality to bound the integral in \eqref{eq:weakRLB} and then rearranging terms leads to 
\begin{align}
   I(X_0; Y_{\sim 0}(t)) \ge   C - \frac{R}{ 1- t}. \label{eq:weakLB_d}
\end{align}

The last step is to relate the mutual information to the overlap matrix. Recalling that $X_0$ has uniform prior distribution $\pr{X_0 = x} = 1/q$ and posterior distribution
\begin{align}
 \pr{X_0 =  x \mid Y_{\sim 0}(t)} = \Psi_x(t),
\end{align}
is the conditional pmf given the output of the coset channel, we can write:
\begin{align}
I(Y_0; Y_{\sim 0}(t)) 
& = \ex{ \sum_{x \in \cX} \Psi_x(t) \log_q q \Psi_x(t)  } \\
& \le \ex{ \log_q q  \sum_{x \in \cX} \Psi^2_x(t)  } \\
& \le  \log_q q  \sum_{x \in \cX} \ex{ \Psi^2_x(t)}  \\
& = \log_q \gtr( Q(t)), 
\end{align}
where the second and third steps are due to Jensen's inequality and the last step follows from the definition of the overlap matrix.  Combining this bound with \eqref{eq:weakLB_d} completes the proof. 
\end{proof}

\begin{proof}[Proof of Lemma~\ref{lem:coset_ser}] 
Set $t^* \coloneqq \frac{1}{2} (1- R/C)$ and observe that,  by the non-negativity of $\delta(t)$, there exists at least one point $s \in [0,t^*]$ such that
\begin{align}
 \dexit(s) \le \frac{1}{t^*} \int_0^{t^*} \dexit(t) \,dt  \le \frac{1}{t^*} \dexit_\mathrm{avg} = \frac{2\dexit_\mathrm{avg}}{1- R/C}. \label{eq:delta_s_bnd}
\end{align}
In the following, we will show that under the assumptions on $\dexit_\mathrm{avg}$, the overlap matrix $Q(s)$ for the channel from $X_0$ to $Y_{\sim 0}(s)$ satisfies
\begin{align}
   q - \gtr(\kexit(s))  \le 2 q\delta(s). 
   \label{eq:delta_s_bnd_a}
\end{align}
The desired bound on the symbol error rate then follows from \eqref{eq:delta_s_bnd},  the bound on the SER in Lemma~\ref{lem:SERtoOverlap} (evaluated with the uniform input distribution), and the fact that $\ser(X_0 \mid Y(t))$ is nondecreasing in $t$. 

To prove \eqref{eq:delta_s_bnd_a}, we first use  Lemma~\ref{lem:Bconstraints} to show that $\gtr(\kexit(s))$ is a distance at most $2 q\delta(s)$ from one of its boundary points (either $1$ or $q$). 
Noting that the symmetry group of a channel is preserved under composition with an erasure channel, we see that the assumptions on the symmetry group of the coset channel (stated for $t=0$) also apply for the channel with parameter $t=s$. Furthermore, in the case that $q$ is prime and the symmetry group of the coset channel is transitive (but necessarily doubly transitive), the assumption
\[
\dexit_\mathrm{avg} < \frac{1- R/C }{8 q^2} = \frac{t^*}{4 q^2}, 
\]
in conjunction with \eqref{eq:delta_s_bnd}, establishes that $4 q^2 \delta(s) < 1 $. Having verified the assumptions of   Lemma~\ref{lem:Bconstraints}, we conclude that
\begin{align}
  \min\Big\{ \gtr(\kexit(s))-1  , q - \gtr(\kexit(s))  \Big \} \le 2 q \delta(s). \label{eq:delta_s_bnd_b}
\end{align} 

Next, we use the area theorem bound in  Lemma~\ref{lem:weak_bound} to show that $\gtr(\kexit(s))$ is bounded away from $1$ in terms a quantity that depends only the gap between $R$ and $C$. Noting that $s \le  t^*  = \frac{1}{2}( 1 - R/C )$ we can apply Lemma~\ref{lem:weak_bound} to obtain
\begin{align}
\gtr(\kexit(s)) \ge q^{ C  -  R/(1-s) } > q^{\frac{1}{2} (C- R) }, \label{eq:delta_s_bnd_c}
\end{align}
where the second inequality follows from 
\[
C - \frac{R}{ 1- s} \ge C - \frac{R}{ 1- t^*}  = \frac{C-R}{ 1 + R/C} > \frac{C-R}{2}. 
\]
Combining \eqref{eq:delta_s_bnd_c} with \eqref{eq:delta_s_bnd} and the assumption
\begin{align}
\dexit_\mathrm{avg} &\le  \frac{(1- R/C) ( q^{\frac{1}{2} (C- R)}-1)}{ q} =   \frac {2 t^*( q^{\frac{1}{2} (C- R)} -1)}{ q},
\end{align}
implies that $ \gtr(\kexit(s)) -1  > 2 q \delta(s)$. Hence, the minimum in  \eqref{eq:delta_s_bnd_b} is attained by the second argument and \eqref{eq:delta_s_bnd_a} holds. 
\end{proof}

\begin{proof}[Proof of Lemma~\ref{lem:rate_cover}] Fix $0 \le t \le 1$. By Lemma~\ref{lem:coset_chan_sym},  $\Psi(t)$ is conditionally independent of $X_{\sim 0}$ given $X_0$, and thus
\begin{align}
 \ex{ \Psi(t)  \mid X_{0}}   = \ex{ \Psi(t)  \mid \bX} = \ex{ \Psi(t) \mid X_{\sim 0}}
\end{align}
where the second equality holds because $X_0 - X_{\sim 0} - \Psi(t)$ is a Markov chain. Accordingly, $\dexit(t)$ can be expressed as
\begin{align}
\delta(t) = \ex{ \| \Psi(t) -\ex{\Psi(t) \mid X_{\sim 0}}\|^2}.
\end{align}
The memoryless property of the channel means that that the entries of $Y_{\sim 0}$ are conditionally independent given $X_{\sim 0}$. Therefore, we can apply the Efron-Stein-Steele inequality in Lemma~\ref{lem:efron_stein} for the index set $[N]\setminus \{0\}$ and the collection of subsets $\cA = \{ B \setminus \{0\} : B \in \cB\}$ to obtain 
 \begin{align}
\dexit(t)
& \le \sum_{A \in \cA} \ex{ \| \Psi(t)   - \ex{ \Psi(t)   \mid X_{\sim 0} ,  Y_{A}(t) } \|^2} \notag \\
& \le \sum_{A \in \cA} \ex{ \| \Psi(t)   - \ex{ \Psi(t)   \mid  Y_{A}(t) } \|^2},  \label{eq:DtSumt}
\end{align}
where the second step is the data processing inequality for MMSE. 

Next, we bound the terms on the RHS of \eqref{eq:DtSumt} in terms of mutual information. For $0 \le t \le 1$ and $A \in \cA$, 
\begin{align*}
Y_{A}(t) - Y_{\sim 0}(t) - X_0  - Y_0,
\end{align*}
forms a Markov chain where $Y_0$ is an observation of $X_0$ through channel $W$. Thus can apply Lemma~\ref{lem:strong_concavity} to obtain
\begin{align}
\MoveEqLeft[1] I(X_0 ; Y_0 \mid Y_{A}(t) ) - I(X_0 ; Y_0 \mid Y_{\sim 0}(t) ) \\
& \ge \frac{\lambda^2_\mathrm{min}}{2 \ln \q} \ex{ \| \ex{ \be_{X_0} \mid Y_{\sim 0}(t)  }   - \ex{\be_{X_0} \mid Y_{A }(t) }   \|^2}\\
&= \frac{\lambda^2_\mathrm{min}}{2 \ln \q} \ex{ \|\Psi(t)   - \ex{ \Psi(t)  \mid Y_{A }(t) }   \|^2}. \label{eq:MI2L1}
\end{align}

Finally, we use the area theorem in Lemma~\ref{lem:area} to bound integrals of the mutual information terms. Setting $B = A \cup \{0\}$ we can write
\begin{align}
\MoveEqLeft 
\int_0^1 I(X_0 ; Y_0 \mid Y_{A}(t) )\,  dt \\
&= \int_0^1 I(X_0 ; Y_0 \mid Y_{B\setminus \{0\} }(t))\,  dt\\
&\overset{(a)}{=} \frac{I(X_B ; Y_B)}{ |B|}\\
& \overset{(b)}{=} \frac{ H(X_B) }{ |B|}  - \frac{ H(X_B  \mid Y_B) }{|B| } \\
& \overset{(c)}{=} \frac{ H(X_B) }{ |B|}  - \int_0^1 H(X_0 \mid Y_{B} , X_{B \setminus \{0\}}(t)) \, dt \\
& \overset{(d)}{\le} \frac{ H(X_B) }{ |B|}  - \int_0^1 H(X_0 \mid \bY  , X_{ \sim 0}(t)) \, dt\\
& \overset{(e)}{=} \frac{ H(X_B) }{ |B|}  - \frac{H(\bX \mid \bY ) }{N}  \label{eq:MIarea_a}
\end{align}
where $(a)$ follows from \eqref{eq:IXY_area}, $(b)$ is the chain rule for entropy, $(c)$ follows from \eqref{eq:HXY_area}, $(d)$ holds because conditioning cannot increase entropy, and $(e)$ is a second application of \eqref{eq:HXY_area}.  Similarly, 
\begin{align}
\int_0^1 I(X_0 ; Y_0 \mid Y_{\sim 0}(t) )\,  dt & = \frac{I(\bX ; \bY )}{ N} \\
& = \frac{ H(\bX) }{ N}  - \frac{ H(\bX  \mid \bY ) }{N }.\label{eq:MIarea_b}
\end{align}
Combining \eqref{eq:MI2L1}, \eqref{eq:MIarea_a}, and \eqref{eq:MIarea_b} yields 
\begin{align}
\MoveEqLeft \frac{\lambda^2_\mathrm{min}}{2 \ln \q}  \int_0^1 \ex{ \| \Psi(t)   - \ex{\Psi(t) \mid Y_{A }(t) }   \|^2} \, dt\\
   &\le  \int_0^1 I(X_0 ; Y_0 \mid Y_{A}(t) ) - I(X_0 ; Y_0 \mid Y_{\sim 0}(t) )  \, dt\\
& \le  \frac{H(X_B )}{ |B|} - \frac{H(\bX)}{N}, 
 \end{align}
Comparing  this inequality with \eqref{eq:DtSumt} and noting that $H(\bX)/N = R$ completes the proof. 
\end{proof}

\begin{proof}[Proof of Lemma~\ref{lem:dexit_GRM}]
First, we choose $k=\lceil \frac{1}{2}\log_q m \rceil $ and note that $k \leq 1 + \frac{1}{2} \log_q m \leq \frac{m}{2}$ for $m \geq q^2$ because $q\geq 2$.
Then, we apply Lemma~\ref{lem:rate_cover} with $\cB = \cup_{i\in [q^{m-k}]} B_i$ where $B_0 = [q^{m-k}]$ and $B_i = [q^m]\setminus \{i\}$ for $i=1,2,\ldots,q^{m-k}$.
By Lemma~\ref{lem:grm_punct}, the punctured code $\cC_{B_0}$ is RM$_q (r,m-k)$ which is transitive and the difference in base-$\q$ entropies equals the difference in code rates under our convention.
For $B_i$ with $1 \leq i < q^{m-k}$, the punctured code $\cC_{B_i}$ equals RM$_q (r,m)$ with one symbol punctured.
The code $\cC_{B_i}$ is transitive because RM$_q (r,m)$ is doubly transitive and this implies
\begin{align*}
\sum_{i=0}^{q^{m-k}-1} & \left( \frac{ H(X_{B_i})}{ |B_i|} - \frac{ H(\bX)}{N}  \right) = R_q (r,m-k) - R_q (r,m) \\ 
+ (&q^{m-k}-1) \left( \frac{q^m R_q (r,m)}{q^m - 1} - R_q (r,m) \right) \\
& \overset{(a)}{\leq} \frac{4 k}{\sqrt{m-k}} + \frac{q^{m-k}-1}{q^m-1} \\
&\overset{(b)}{\leq} \frac{4 \sqrt{2} k}{\sqrt{m}} + \frac{1}{q^k} \\
& \overset{(c)}{\leq} \frac{4 \sqrt{2} (1+\frac{1}{2}\log_q m ) }{\sqrt{m}} + \frac{1}{\sqrt{m}},
\end{align*}
where $(a)$ follows from Lemma~\ref{lem:grm_rate_diff}, $(b)$ holds because $k \leq \frac{m}{2}$, and $(c)$ follows from $k=\frac{1}{2}\lceil \log_q m \rceil \leq 1+ \frac{1}{2} \log_q m$.
Upper bounding the constants by integers gives the final result.
\end{proof}

\section{Auxiliary Results}

\begin{lemma}[Efron-Stein-Steele]\label{lem:efron_stein}
Let $\bY \in \cY^N$ be a random vector
whose entries are conditionally independent given sigma-algebra $\cG$. Let $Z$ be a measurable function of $\bY$ taking values in a finite-dimensional vector space with $\ex{ \|Z\|^2} < \infty$. For any collection $\cA$ of subsets of $[N]$ with $\bigcap_{A \in \cA} A = \emptyset$, 
\begin{align*}
\ex{ \|Z- \ex{Z \mid \cG}\|^2} \le \sum_{A \in \cA} \ex{ \| Z  - \ex{ Z \mid \cG,  Y_A}\|^2}.
\end{align*}
\end{lemma}
\begin{proof}
Set $K = |\cA|$ and let $\{A_1, \dots, A_{K}\}$ be an enumeration of the sets in $\cA$. Let $S_0 \supseteq S_1 \supseteq \dots \supseteq S_K$ be the nonincreasing sequence of sets by defined by $S_0 = [N]$ and $S_k = S_{k-1} \cap  A_k$ for $k = 1, \dots, K$.  By assumption,  $S_K = \emptyset$. 

Let $\cF_k$ be the sigma-algebra generated by $(\cG, Y_{S_k})$. Noting that $Z = \ex{ Z \mid \cF_0}$ and $\ex{ Z\mid \cG} = \ex{ Z \mid \cF_K}$, we obtain the orthogonal decomposition:
\begin{align} \label{eq:orthdecomp}
\bEx[\|Z  - \ex{Z | \cG}\|^2] 
&= \sum_{k=1}^{K} \bEx \| \ex{ Z| \cF_{k-1}}     - \ex{ Z| \cF_{k}} \|^2.   \qquad
\end{align}
By assumption, the entries of $\bY$ are conditionally independent given $\cG$. Recalling that $S_k = S_{k-1} \cap A_k$, it follows that 
\begin{align}
\ex{ \Psi \mid \cF_k} 
& =  \ex{\ex{ \Psi  \mid \cG, Y_{A_k } } \mid \cF_k}\\
& =  \ex{ \ex{ \Psi  \mid \cG, Y_{A_k } } \mid \cF_k, Y_{ S_{k-1}\setminus A_k }}\\
& =   \ex{ \, \ex{ \Psi \mid \cG, Y_{A_k} }  \mid \cF_{k-1}}, 
\end{align} 
where the first equality is the tower property of conditional expectation and the second step holds because $\ex{ \Psi  \mid \cG, Y_{A_k }}$ and $Y_{ S_{k-1}\setminus A_k }$ are conditionally independent given $\cF_k$. 

Hence, 
 \begin{align*}
\MoveEqLeft  \ex{ \|  \ex{ \Psi \mid  \cF_{k-1}}     - \ex{ \Psi \mid \cF_{k}} \|^2}  \\
& =   \ex{ \|  \ex{ Z \mid \cF_{k-1}}  - \ex{ \ex{ Z \mid  \cG, Y_{A_k} }  \mid  \cF_{k-1}}    \|^2}  \\
 & \le  \ex{ \| Z - \ex{ Z \mid  \cG, Y_{A_k}  } \|^2},
 \end{align*}
where the second step is Jensen's inequality. 
Plugging this inequality back in \eqref{eq:orthdecomp} and  noting that the bound does not depend on the enumeration of $\cA$ completes the proof. 
\end{proof}

The following result is closely related to the extrinsic information transfer (EXIT) area theorem~\cite[Thm.~1]{Ashikhmin-it04} though applied to transitive codes and written in our notation.

\begin{lemma}[EXIT Area Theorem] \label{lem:area}
Let $\bY \in \cY^N$ be an observation of input $\bX \in \cX^N$ though through a memoryless channel $W$. If $S \subseteq [N]$ is a subset that contains $0$ and the symmetry group of $X_S$ is transitive, then
\begin{align}
\frac{I(X_S ; Y_S)}{ |S|}  &= \int_0^1 I(X_0; Y_0 \mid Y_{S \setminus \{0\}}(t))\, dt \label{eq:IXY_area}\\
\frac{H(X_S \mid  Y_S)}{ |S|}  &= \int_0^1 H(X_0\mid Y_S,  X_{S \setminus \{0\}}(t))\, dt. \label{eq:HXY_area}
\end{align}
where $\bX(t)$  and $\bY(t)$ are observations of $\bX$ and $\bY$, respectively,  through a memoryless erasure channel with erasure probability $t \in [0,1]$. 
\end{lemma}

\begin{proof}
Without loss of generality we may assume $S = [N]$.  By the fundamental theorem of calculus,  
\begin{align*}
I(\bX ; \bY) & = H(\bX \mid \bY(1)) - H(\bX \mid \bY(0))  \\
& = \int_0^1 \frac{d}{dt} H(\bX \mid \bY(t))\, dt.
\end{align*}
To proceed, we use the proof technique from~\cite{Measson-it08} and consider the setting where the $i$-th erasure channel has erasure probability $t_i$. Writing $\bt = (t_i)_{i \in [N]}$, the erasure channel output becomes $\bY(\bt)$ and the law of the total derivative gives
\begin{align*}
\frac{d}{dt} H(\bX\mid \bY(t) ) = \sum_{i \in S} \partial_i H( \bX \mid  \bY(\bt) ) \Big \vert_{t_i = t, \forall i \in [N] },
\end{align*}
where $\partial_i$ is the partial derivative operator with respect to the $i$-th coordinate of $\bt$. 
By the chain rule for entropy and  that fact that $Y_i(t_i) \to X_i \to Y_{\sim i}(t_{\sim i} )$ is a Markov chain (because the channel is memoryless), the entropy can be decomposed as
\begin{align*}
H(\bX  \mid \bY(\bt) )
& = H(X_{\sim i}  \mid X_i,  Y_{\sim i}(t_{\sim i}  )) \\
& \quad + (1- t_i) H(X_i \mid Y_i, Y_{\sim i}(t_{\sim i}) )\\
&\quad + t_i  H(X_i \mid  Y_{\sim i}(t_{\sim i} ) ).
\end{align*}
Thus, the $t_i$-derivative is equal to $I(X_i ; Y_i \mid   Y_{\sim i}(t_{\sim i}))$.  If the permutation automorphism of the code is transitive, then the derivative is the same for all $i \in [N]$.
Hence, $\frac{d}{dt} H(\bX \mid \bY (t) ) = N \, I(X_0 ; Y_0 \mid   Y_{\sim 0}(t))$ and~\eqref{eq:IXY_area} follows.

The integral expression for $H(\bX \mid \bY)$ follows similarly and the proof is omitted. 
\end{proof}

\fi 

\end{document}